\documentclass[fontsize=10pt,a4paper]{article}

\usepackage{graphicx}
\usepackage{amsmath}
\usepackage{amssymb}
\usepackage{amsthm}
\usepackage{color}
\usepackage{natbib}

\usepackage[utf8]{inputenc}
\usepackage[T1]{fontenc}

\title{\textbf{Pure single photon generation by type-I\\ PDC with backward-wave amplification}}

\date{}
\author{\small{A. Christ, A. Eckstein, P. J. Mosley and, C. Silberhorn}}

\begin{document}
\maketitle
\vspace{-1cm}
\begin{center}
	\textit{\small{Max-Planck Institute for the Science of Light\\ G\"unther Scharowsky-Str.1 / Bau 24, 91054 Erlangen, Germany}}
\end{center}


\begin{abstract}
We explore a promising method of generating pure heralded single photons. Our
approach is based on parametric downconversion in a periodically-poled
waveguide. However, unlike conventional downconversion sources, the photon pairs
are counter-propagating: one travels with the pump beam in the forward direction
while the other is backpropagating towards the laser source. Our calculations
reveal that these downconverted two-photon states carry minimal spectral
correlations within each photon-pair. This approach offers the possibility to
employ a new range of downconversion processes and materials like PPLN
(previously considered unsuitable due to its unfavorable phasematching
properties) to produce heralded pure single photons over a broad frequency range.
\end{abstract}

\normalsize
\section{Introduction}
Linear optical quantum computing (LOQC) schemes, such as continuous 
variable entanglement distillation \cite{braunstein_quantum_2005} or single photon 
quantum gates \cite{Ralph01} require sources of pure 
heralded single photons. Such single photon sources may be realized via  
photon pair generation by parametric downconversion (PDC).
The photon number correlation between the resulting fields, typically labelled
signal and idler, can be exploited to herald the existence of one photon by
detection of its partner. However the purity of the heralded photon is limited
by spatial and spectral correlations within each photon pair arising from energy and momentum conservation between pump, signal and idler photons.
One possibility of generating pure heralded single photons without spectral filtering and a reduction in the source brightness is group velocity matching
 \cite{mosley:133601}.
However this approach to generate separable photon pairs is limited to a few materials and wavelength ranges.

Most PDC experiments to date have been performed in bulk crystals, yet lately a
lot of attention has been focused on PDC in waveguides. The main advantage of
PDC in rectangular waveguides is the strict collinear propagation of the
pump, signal and idler fields, in contrast to angular dispersion in bulk crystal setups. Along with the high modal confinement inside the waveguide this leads to a large increase in collection efficiency \cite{Fiorentino:07} and the elimination of spatial correlations. Furthermore, due to the strict collinear propagation of pump, signal and idler beams these sources are much more convenient to handle in the laboratory. 

The spectral properties of PDC states are governed by the phasematching properties of the nonlinear material and this determines the frequencies of the downconverted photons. In a bulk nonlinear crystal one can exploit noncollinear PDC to achieve perfect phasematching, but this approach cannot be used in a waveguide structure as the direction of propagation is fixed. Instead, one must adopt quasi-phasematching (QPM) \cite{PhysRevA.66.013801,hum_quasi-phasematching_2007}: A spatial periodic variation of the \(\chi_{ijk}^{\left(2\right)}\)-nonlinearity in the crystal introduces a new so called quasiphasematching vector (\(k_{QPM} = 2 \pi / \Lambda\)), for a sinusoidal poling with period \(\Lambda\). In that way it is in principle possible to choose the signal and idler wavelengths freely, under the restriction of energy conservation. 

The generation of backward-wave oscillations in three-wave-mixing processes was proposed in 1966 \cite{harris:114}. For the generation of correlated photon pairs in waveguided PDC this approach was revisited in 2002 by Booth et. al. \cite{booth-2002-66}. 
In this configuration almost all the momentum of the pump photon has to be
compensated by the QPM poling structure within the crystal. This requires
grating periods in the sub-micron range (0.2 -0.6 \(\mu\)m) for signal and idler
photons generated in the range from 800 to 1600 nm. 

Quasiphasematched PDC processes with counterpropagating signal and idler
photons and a perpendicularly propagating pump in planar semiconductor waveguides have already been observed \cite{lanco:173901,ravaro_nonlinear_2005}, and their respective quantum properties have been studied \cite{jr.:013803}. However, to date, the high absorption in semiconductor materials and tiny interaction volume limit the achievable photon flux. In dielectric materials a breakthrough has been made towards sub-micron poling periods in the past year: Backward-wave  oscillation in potassium titanyl phosphate (KTP) has been reported \cite{canalias_mirrorless_2007}, and simultaneously in lithium niobate (LN) new sub-micron poling techniques have been explored \cite{minakata_nanometer_2006}. Therefore the required crystals are within reach.

In this paper, we consider a collinear waveguided PDC setup with 
counterpropagating signal and idler fields. We show that this configuration allows 
the generation of pure heralded single photons in a large range of nonlinear 
materials and wavelengths.

\section{The PDC state}

The frequency structure of a downconverted  \textit{copropagating} two-photon state is found to be \cite{Grice97}: 
\begin{multline}
	\left|\psi_{s,i}\right\rangle \approx \left|0\right> + A \int\!\!\int\!\textrm{d} \omega_s \textrm{d}\omega_i 
	\underbrace{\exp\left[-\frac{\left(\omega_s + \omega_i -
    \omega_p\right)^2}{2
    \sigma^2}\right]}_{\alpha\left(\omega_s+\omega_i\right)}\\
    \times \underbrace{\rm{sinc}\left(\frac{L}{2}\Delta k\right) \exp\left[- \imath \frac{L}{2} \Delta k \right]}_{\phi\left(\omega_s,\omega_i\right)}
\hat{\textrm{a}}^\dagger_s(\omega_s) \hat{\textrm{a}}^\dagger_i(\omega_i) \left|0\right> 
\label{pdc-state}
\end{multline}
The pump distribution \(\alpha(\omega_s+\omega_i)\) is given by the incoming laser. In our case we assume a mode-locked pulsed laser system with a Gaussian frequency distribution, centered around \(\omega_p\) with width \(\sigma\). The phasematching function \(\phi(\omega_s,\omega_i)\) is governed by the waveguide dimensions and crystal dispersion, ensuring momentum conservation (\(\Delta k = k_p - k_s - k_i -k_{QPM}\)). Because of the strict collinear propagation inside the waveguide the transverse wavevector mismatch does not enter Eq. (\ref{pdc-state}). The product of these two functions gives the joint spectral amplitude (JSA): \(f(\omega_s,\omega_i)=\alpha(\omega_s+\omega_i)\cdot\phi(\omega_s,\omega_i)\). 
For analytic calculations the two-photon state is often simplified with the Gaussian approximation (\(\rm{sinc}(x) \approx \exp(-\gamma x^2)\) where \(\gamma \approx 0.193\)).

Heralding one photon of this PDC-state will in general lead to a mixed
heralded single photon state, due to correlations in the JSA (\(f\left(\omega_s,
\omega_i\right)\)) \cite{uren-2005-15}.
Pure heralded single photons are created if and only if the downconverted two-photon state can be written as a product state \(\left|\psi_{s,i}\right\rangle = \left|\psi_{s}\right\rangle \otimes \left|\psi_{i}\right\rangle\).
This requires a separable JSA: \(f(\omega_s,\omega_i)=f(\omega_s)f(\omega_i)\).
In order to quantify the separability of the generated PDC states one has to
perform a Schmidt decomposition \cite{law_continuous_2000}, i.e. a basis
transformation into a set of orthonormal Schmidt modes,
\(\left|\psi_s^n(\omega_s)\right\rangle\) and \(\left|\psi_i^n(\omega_i)\right\rangle\): 
\begin{equation}
	\left|\psi_{s,i}(\omega_s,\omega_i)\right\rangle = \sum_n \sqrt{\lambda_n} \left|\psi_s^n(\omega_s)\right\rangle \otimes \left|\psi_i^n(\omega_i)\right\rangle
	\label{schmidt-decomposition}
\end{equation}
The probability to emit a photon pair into one specific pair of Schmidt modes \(  \left|\psi_s^n(\omega_s)\right\rangle \otimes \left|\psi_i^n(\omega_i)\right\rangle\) is given by \(\lambda_n\), which is monotonically decreasing for successive higher order modes. Thus a perfectly separable state corresponds to a state where the full weight of the probability distribution accumulates on the first pair of Schmidt modes (\(\lambda_0 =1\)). 

\section{Backward-wave oscillations}
\begin{figure}[htb]
		\centering\includegraphics[width=\textwidth]{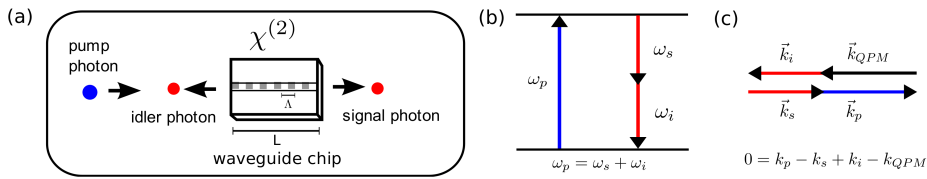}
		\caption{Waveguided parametric downconversion with one backward-wave oscillation: (a) Process scheme, (b) Energy conservation, (c) Momentum conservation.}
		\label{counterpropagation}
\end{figure} 
Fig. \ref{counterpropagation} illustrates the generation of
\textit{counterpropagating}
photon pairs and the corresponding energy and momentum conservation conditions.
To account for the backward propagating wave in our formalism
(Eq.(\ref{pdc-state})),  we have to alter the momentum conservation condition: 
\begin{equation}
	\Delta k = k_p - k_s \textcolor{red}{+} k_i - 2 \pi /\Lambda.
	\label{alteredMomentumConservation}
\end{equation}
This has a significant effect on the properties of the generated
two-photon-states. The biggest advantage over copropagating PDC are the
different requirements to generate separable two-photon states. With the
Gaussian approximation and expansion of the wave vector mismatch \(\Delta k\) as
a Taylor series up to the first order, we are able to analytically derive a
condition for separability from Eq. (\ref{pdc-state}) and (\ref{alteredMomentumConservation}) (analogous to \cite{uren-2005-15}):
\begin{equation}
	0 = \frac{2}{\sigma^2} + \left(k_p' - k_s'\right) \left(k_p' \textcolor{red}{+} k_i'\right) = 
	\frac{2}{\sigma^2} + \left(\frac{1}{v_p} - \frac{1}{v_s}\right) \left(\frac{1}{v_p}\textcolor{red}{+} \frac{1}{v_i}\right). 
	\label{decorrelation condition}
\end{equation}
Therefore a waveguide in which the pump pulses propagate faster than the downconverted forward propagating signal pulses (\(v_s < v_p\)) will generate separable photon pairs. Note that this requirement is much easier to satisfy than that for the usual copropagating case, where the group velocities must satisfy either  \( v_s < v_p < v_i\) or \(v_i < v_p < v_s\) \cite{uren-2005-15}.

Further insight is obtained by deriving the angle of the phasematching function in the \(\left\{\omega_s, \omega_i\right\}\)-plane:
\begin{equation}
	\theta = - \rm{arctan}\left[\frac{k_p' - k_s'}{k_p' \textcolor{red}{+} k_i'}\right] 
	= - \rm{arctan}\left[\frac{\nu_{s} - \nu_p}{\nu_{i} \textcolor{red}{+} \nu_p}  \frac{\nu_i}{\nu_s} \right] 
	\label{phasematching-angle}
\end{equation}
Here \(\theta\) is defined as the angle between the phasematching function and the signal axis. As can be deduced from Eq. (\ref{decorrelation condition}) and  (\ref{phasematching-angle}), the condition for factorability requires a phasematching angle between \(0^\circ < \theta < 90^\circ\). Considering the relative magnitude of the numerator and denominator in Eq.(\ref{phasematching-angle}), identical group velocities for the signal, idler and pump waves will result in a horizontally orientated phasematching function and a factorable JSA. This is in very good agreement with the small group velocity dispersion in common nonlinear materials over the relevant wavelength region \(\nu_s(\omega_s) \approx \nu_i(\omega_i) \approx \nu_p(\omega_s+\omega_i)\). These requirements are  opposed to copropagating decorrelation proposals which rely on different group velocities for the signal, idler and pump waves. In the counterpropagating case very similar group velocities are demanded, a requirement that is much more robust and easier to fulfill. 

\begin{figure}
	\centering\includegraphics[width=\textwidth]{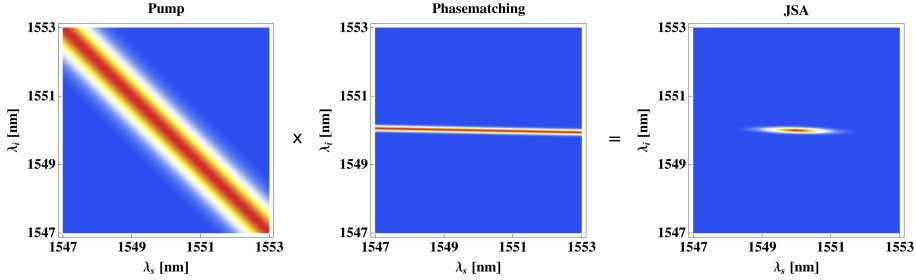}
		\caption{Pump envelope, Phasematching function and JSA plotted in the
        Gaussian approximation without phase; Parameters: LN, type-I PDC,
        e-polarized rays, pump central wavelength \(\lambda_p\)  = 775 nm, FWHM
        of the pump intensity distribution  \(\Delta \lambda_p  = 0.58\) nm,
        waveguide dimensions: 4 \(\mu\)m x  4 \(\mu\)m x \(5\) mm, grating period
        \(\Lambda= 0.35\) \(\mu\)m.}
		\label{counter:picture}
\end{figure}

The highly nonlinear crystal LN, commonly used in PDC experiments to generate photon pairs is unable to produce factorable copropagating photon pairs, but can be used to generate separable pairs in the backward-propagating regime.
Fig. \ref{counter:picture} illustrates this particular example of a separable counterpropagating two-photon state. It demonstrates several benefits of the backward-wave approach simulated using periodically poled LN (PPLN) as the nonlinear medium. The momentum mismatch changes to a much stricter condition in comparison with copropagating PDC. This leads to an extremely narrow spectral width of the backpropagating photons, almost one order of magnitude smaller than the spectral distribution of the forward propagating photon. In this case, the FWHM of the wavelength distribution is 0.09 nm for the backpropagating photon and 0.73 nm for the forward propagating photon. The narrow frequency bandwidth of the backpropagating photon makes it very well suited for long distance transmission in optical fibers. 

It has to be noted that according to \cite{PhysRevA.70.043810}, the total photon
pair production rate will decrease with respect to sources that create strictly
forward propagating photons, but the effective generation rate can still exceed
bulk crystal setups due to the high nonlinearity and higher collection efficiencies in a waveguide architecture.

In more general terms we would like to emphasise that the separability of a
downconverted photon pair is almost independent of the signal and idler
frequencies (Fig. \ref{decorrelation:tuning}(a)). With appropriate grating
periods it is possible to generate  separable degenerate signal and
idler photons from 800 nm (where detectors are most efficient) to 1550 nm (the
wavelength with minimal loss in optical fibers). This can be used to create a \textit{tuneable} pure heralded single photon source. As depicted in Fig. \ref{decorrelation:tuning}(b), for a fixed grating period different pump wavelengths lead to a change in the downconverted signal wavelength, whereas the separability and the idler frequency remain constant.  Hence the wavelength of the signal photon can be tuned by changing the pump wavelength, without impact on the idler frequency and with very little change in the separability.

\section{Numerical analysis}
To quantify the stated benefits, we numerically investigated periodically poled LN (PPLN) and periodically poled KTP (PPKTP) as sources of separable counterpropagating photon pairs. 
In PPLN we chose the type-I downconversion process with the highest
nonlinearity, where extraordinary-polarized pump photons decay into
extraordinary-polarized signal and idler photons (\(\chi^{\left(2\right)}_{eee}=
63\) pm/V). 
 In KTP we analyzed strictly z-polarized signal, idler and pump waves making use
 of the largest tensor element \(\chi^{\left(2\right)}_{zzz}= 27.4\) pm/V. 

The rectangular waveguide embedded in the crystal material was modelled  with
standard dimensions of 4 \(\mu\)m x 4 \(\mu\)m x 5 mm  and a realistic refractive
index step between waveguide and surrounding material of 0.01 
. To simulate the propagation of our signal, idler and pump
waves inside the waveguide we calculated the spatial modes of signal, idler and
pump fields according to \cite{marcuse_theory_1974, toBePublished}. The resulting
decomposition of the wavevector into its longitudinal and transverse components
enabled us to correct the bulk crystal Sellmeier equations and to obtain a
modified JSA. In the scope of this paper we assume that the signal, idler and pump
fields propagate in the fundamental waveguide mode.

\begin{figure}[!ht]
	\centering
	\includegraphics[width=\textwidth]{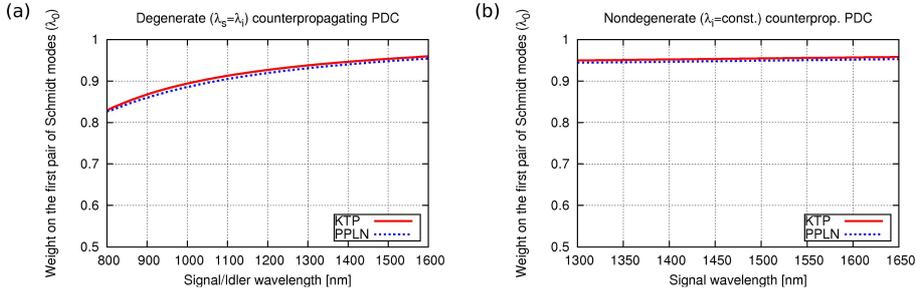}
	\caption{(a) In counterpropagating PDC it is possible to generate separable degenerate signal and idler photons in the range from 800 to 1600 nm. (b) The separability is maintained for nondegenerate PDC (\(\lambda_i =1550\) nm).}
	\label{decorrelation:tuning}
\end{figure} 

We investigated the possibility of generating decorrelated and degenerate photon
pairs in the range of 800 nm to 1600 nm. For each studied signal and idler
degeneracy wavelength the pump wavelength was chosen appropriately (\(\lambda_p
= \lambda_{s,i}/2\)), and the grating was matched to give phasematching at the
degeneracy point (\(\Lambda = 2 \pi / (k_p-k_i-k_s)\)).  For each parameter set
(\(\lambda_s =  \lambda_i = 2 \lambda_p\), \(\Lambda(\lambda_s,\lambda_i)\)) we optimized the pump width \(\Delta
\lambda_p\) in the range from  \(\Delta \lambda_p = 0.02-0.35\) nm to yield a state
with  maximum  separability. The optimum value of \(\Delta \lambda_p\) was
determined by performing a Schmidt decomposition according to  
Eq. (\ref{schmidt-decomposition}) for every set of parameters. 

The results are depicted in Fig. \ref{decorrelation:tuning}(a): The overall
probability of the generated photon pairs to be emitted in the first pair of
Schmidt modes (\(\lambda_0\)) is very high and only differs slightly for KTP and PPLN. The general improvement in separability for higher wavelengths is due to the fact that in this region the group velocities of the signal, idler and pump waves almost perfectly equal each other.

In a similar manner we checked the feasibility of this setup as a frequency-tuneable pure heralded single photon source using nondegenerate PDC (Fig. \ref{decorrelation:tuning}(b)). For a constant grating period \(\Lambda\) yielding phasematching at 1550 nm for signal and idler, we investigated the impacts of  tuning  the pump central frequency. Due to the horizontal orientation of the phasematching function the frequency distribution of the idler photons remains unchanged. However the frequency of the forward propagating signal photon shifts with the pump frequency.
Again the pump width was chosen to yield a maximally separable two-photon states,
now in the range from \(\Delta \lambda_p = 0.22 - 0.34\) nm. Once more our results are almost independent of the chosen nonlinear crystal and the constant high level of separability shows that with this setup it will indeed be possible to create a tuneable pure heralded single photon source.

\section{Conclusion}
We have examined the spectral properties of downconverted counterpropagating two-photon states in rectangular waveguides. The major differences in comparison with copropagating downconverted photon-pairs allow us to exploit a wide range of processes and materials for heralded single photon generation. This technique provides separable two-photon states for a wide range of degenerate and non-degenerate signal and idler wavelengths which will be useful for practical purposes such as LOQC. 
Due to the progress in the production of microstructured waveguides an experimental implementation of our proposal will be feasible in the near future. 

The authors would like to thank Wolfgang Mauerer for his support on the
numerical analysis.

\bibliographystyle{abbrv}

\begin{thebibliography}{99}
\setlength{\itemsep}{0pt}
\setlength{\parsep}{0pt}
\setlength{\parskip}{0pt}


\bibitem{braunstein_quantum_2005}
S.~L. Braunstein and P.~van Loock, ``Quantum information with continuous
  variables,'' Rev. Mod. Phys. \textbf{77}, 513--577 (2005).

\bibitem{Ralph01}
T.~C. Ralph, A.~G. White, W.~J. Munro, and G.~J. Milburn, ``Simple
  scheme for efficient linear optics quantum gates,'' Phys. Rev. A
  \textbf{65}, 012314--012320 (2001).

\bibitem{mosley:133601}
P.~J. Mosley, J.~S. Lundeen, B.~J. Smith, P.~Wasylczyk, A.~B. U'Ren,
  C.~Silberhorn, and I.~A. Walmsley, ``Heralded Generation of Ultrafast
  Single Photons in Pure Quantum States,'' Phys. Rev. Lett.
  \textbf{100}, 133601--133605 (2008).

\bibitem{Fiorentino:07}
M.~Fiorentino, S.~M. Spillane, R.~G. Beausoleil, T.~D. Roberts, P.~Battle, and
  M.~W. Munro, ``Spontaneous parametric down-conversion in periodically
  poled KTP waveguides and bulk crystals,'' Opt. Express \textbf{15},
  7479--7488 (2007).

\bibitem{PhysRevA.66.013801}
G.~Di~Giuseppe, M.~Atat\"ure, M.~D. Shaw, A.~V. Sergienko, B.~E.~A. Saleh, and
  M.~C. Teich, ``Entangled-photon generation from parametric
  down-conversion in media with inhomogeneous nonlinearity,'' Phys. Rev. A
  \textbf{66}, 013801--013818 (2002).

\bibitem{hum_quasi-phasematching_2007}
D.~S. Hum and M.~M. Fejer, ``Quasi-phasematching,'' Comptes Rendus
  Physique \textbf{8}, 180--198 (2007).

\bibitem{Grice97}
W.~P. Grice and I.~A. Walmsley, ``Spectral information and
  distinguishability in type-II down-conversion with a broadband pump,'' Phys.
  Rev. A \textbf{56}, 1627--1634 (1997).

\bibitem{uren-2005-15}
A.~B. U'Ren, C.~Silberhorn, R.~Erdmann, K.~Banaszek, W.~P. Grice, I.~A.
  Walmsley, and M.~G. Raymer, ``Generation of Pure-State Single-Photon
  Wavepackets by Conditional Preparation Based on Spontaneous Parametric
  Downconversion,'' Laser Phys. Lett. \textbf{15}, 146--160 (2005).

\bibitem{law_continuous_2000}
C.~K. Law, I.~A. Walmsley, and J.~H. Eberly, ``Continuous Frequency
  Entanglement: Effective Finite Hilbert Space and Entropy Control,'' Phys.
  Rev. Lett. \textbf{84}, 5304--5307 (2000).

\bibitem{harris:114}
S.~E. Harris, ``Proposed backward wave oscillation in the infrared,''
  Appl. Phys. Lett. \textbf{9}, 114--116 (1966).

\bibitem{booth-2002-66}
M.~C. Booth, M.~Atature, G.~{Di Giuseppe}, A.~V. Sergienko, B.~E.~A. Saleh, and
  M.~C. Teich, ``Counter-propagating entangled photons from a waveguide
  with periodic nonlinearity,'' Phys. Rev. A \textbf{66}, 023815--02323 (2002).

\bibitem{lanco:173901}
L.~Lanco, S.~Ducci, J.-P. Likforman, X.~Marcadet, J.~A.~W. van Houwelingen,
  H.~Zbinden, G.~Leo, and V.~Berger, ``Semiconductor Waveguide Source of
  Counterpropagating Twin Photons,'' Phys. Rev. Lett. \textbf{97},
  173901--173905 (2006).

\bibitem{ravaro_nonlinear_2005}
M.~Ravaro, Y.~Seurin, S.~Ducci, G.~Leo, V.~Berger, A.~de~Rossi, and G.~Assanto,
  ``Nonlinear AlGaAs waveguide for the generation of counterpropagating
  twin photons in the telecom range,'' J. Appl. Phys. \textbf{98},
  063103--063109 (2005).

\bibitem{jr.:013803}
J.~P. Jr., ``Quantum properties of counterpropagating two-photon states
  generated in a planar waveguide,'' Phys. Rev. A 
   \textbf{77}, 013803--013817 (2008).

\bibitem{canalias_mirrorless_2007}
C.~Canalias and V.~Pasiskevicius, ``Mirrorless optical parametric
  oscillator,'' Nat Photon \textbf{1}, 459--462 (2007).

\bibitem{minakata_nanometer_2006}
M.~Minakata, M.~Islam, S.~Nagano, S.~Yoneyama, T.~Sugiyama, and H.~Awano,
  ``Nanometer size periodic domain inversion in LiNbO3 substrate using
  circular form full cover electrodes,'' Solid-State Electron. \textbf{50},
  848--852 (2006).

\bibitem{PhysRevA.70.043810}
A.~N. Vamivakas, B.~E.~A. Saleh, A.~V. Sergienko, and M.~C. Teich,
  ``Theory of spontaneous parametric down-conversion from photonic
  crystals,'' Phys. Rev. A \textbf{70}, 043810-043817 (2004).

\bibitem{marcuse_theory_1974}
D.~Marcuse, ``Theory of dielectric optical waveguides'', 49--59   (1974).

\bibitem{toBePublished}
 A. Christ, A. Eckstein, K. Laiho, T. Lauckner, P. J. Mosley, C. Silberhorn,
  ``Spatial to spectral mode mapping in waveguided PDC,'' to be published, (2009).

\end{thebibliography}
\small{

\end{document}